# HeartBERT: A Self-Supervised ECG Embedding Model for Efficient and Effective Medical Signal Analysis

Saedeh Tahery [a, †], Fatemeh Hamid Akhlaghi [b, a, †], Termeh Amirsoleimani [a]

[a] *Faculty of Computer Engineering, K. N. Toosi University of Technology, Tehran, Iran*
[b] *School of Electrical and Computer Engineering, University of Tehran, Tehran, Iran*

This paper is currently under investigation/review, and further changes may be applied in the future.

**Abstract**

The HeartBert model is introduced with three primary objectives: reducing the need for labeled data, minimizing computational resources, and simultaneously improving performance in machine learning systems that analyze Electrocardiogram (ECG) signals. Inspired by Bidirectional Encoder Representations from Transformers (BERT) in natural language processing and enhanced with a self-supervised learning approach, the HeartBert model—built on the RoBERTa architecture—generates sophisticated embeddings tailored for ECG-based projects in the medical domain. To demonstrate the versatility, generalizability, and efficiency of the proposed model, two key downstream tasks have been selected: sleep stage detection and heartbeat classification. HeartBERT-based systems, utilizing bidirectional LSTM heads, are designed to address complex challenges. A series of practical experiments have been conducted to reveal the superiority and advancements of HeartBERT, particularly in terms of its ability to perform well with smaller training datasets, reduced learning parameters, and effective performance compared to rival models. The code and data are publicly available at https://anonymous.4open.science/r/HeartBert.

## 1. Introduction

Biological signals such as Electrocardiograms (ECGs) are a cornerstone diagnostic tool in healthcare, specifically in cardiology, providing a wealth of information about heart function by measuring electrical activity. Beyond basic heart rate assessment, ECGs excel at detecting arrhythmias and irregular heartbeats that can signal underlying issues [1-4]. ECGs can identify signs of a heart attack, ischemia, and structural abnormalities in the heart muscles. Additionally, ECG technology is essential in wearable devices, extending its use beyond hospitals. Wearable ECG devices continuously monitor heart activity outside clinical settings, providing valuable insights into health conditions and aiding in the early detection of abnormalities [5-8]. Using ECG signals is very cost-effective, especially when integrated into wearable devices like smartwatches. ECG signals help with early detection of heart diseases and continuous monitoring of heart activities, allowing for timely treatment and preventing serious health issues. Besides, for individuals suffering from chronic diseases, regular ECG monitoring at home reduces the need for frequent doctor visits, providing convenience and cutting down healthcare costs for both patients and healthcare systems. These applications highlight the significance of ECGs in improving patient outcomes across a spectrum of medical conditions.

Existing machine learning methods for interpreting ECG signals often rely on supervised learning approaches [9], which require labeled datasets and may not adapt well to diverse patient populations or evolving medical scenarios. Collecting labeled datasets poses significant challenges and costs in machine learning, particularly within healthcare settings, where labeling can be prohibitively expensive and grappling with privacy can be dreadful [10]. On the other hand, utilizing unlabeled data can reveal the potential value of unsupervised learning techniques, providing an opportunity to enhance ECG analysis by harnessing the abundance of unlabeled data, thereby reducing dependence on costly labeled datasets and fostering adaptability to varied clinical scenarios. However, the volume of unlabeled data often far surpasses that of labeled data, highlighting the potential of self-supervised representation learning [11-16]. This approach can strengthen model accuracy and robustness even with limited labeled data, a significant advantage in the medical field.

In this paper, we investigate self-supervised representation learning inspired by the Robustly Optimized BERT Approach (RoBERTa) model [17] for text data in the context of ECG signals. By developing a model that learns contextual embeddings from unlabeled ECG signals, we can create a versatile model applicable to diverse downstream tasks in ECG analysis.

To bridge the gap between ECG analysis and natural language processing (NLP), we leverage the inherent similarities in these domains, both involving the analysis of periodic time series data characterized by meaningful patterns. The similar attributes between sentences in natural languages and ECG signals motivate us to develop HeartBERT, a model based on RoBERTa architecture—an encoder-only model—using an innovative method to translate ECG signals into an intermediate synthetic language, aimed at advancing text tokenization and embedding through a transformer-based architecture.

The pre-training process involves porting ECG signals into textual representations through signal quantization and discretization [18, 19], followed by encoding these representations into a language-like format. This transformation allows us to input the derived text data directly into the HeartBERT model for pre-training from scratch in a self-supervised manner. To this end, we first construct a customized tokenizer based on the Byte Pair Encoding [20] that captures the most frequent and meaningful patterns, thereby providing adequate coverage of subword units for downstream tasks. The tokenized data is then processed through transformer layers during pre-training, enabling the HeartBERT model to learn contextual embeddings that encapsulate the underlying patterns and semantics of the given ECG signals. By exclusively training on this synthetic language derived from unlabeled ECG signals, our approach facilitates the creation of an adaptive model capable of extracting and comprehending ECG-specific features. Having completed the pre-training step, the model could be used to fine-tune downstream tasks with limited labeled data utilizing the contextual embeddings provided by the HeartBERT model.

In this paper, we address two clinically significant downstream tasks: sleep-stage classification and heartbeat classification, both of which are crucial for advancing the understanding and management of cardiovascular health. Sleep-stage classification plays a vital role in diagnosing and managing sleep disorders, which are closely linked to conditions such as high blood pressure, diabetes, heart disease, and psychiatric disorders [21-26]. Sleep disorders often lead to daytime sleepiness, reduced energy, and impaired immune function, making sleep studies increasingly important in modern medicine. For the sleep-stage classification task, we conduct experiments on the MIT-BIH Polysomnographic Database [27]. Our results show that utilizing our HeartBERT model leads to significant improvements over baseline models, including widely used deep-learning architectures such as Convolutional Neural Networks (CNNs) [28, 29] and Long Short Term Memory (LSTM) networks [30]. Following a thorough analysis of the effect of fine-tuning the transformer layers of the HeartBERT model, our proposed model achieves the best performance with an F1 score of approximately 75% for the three-stage classification and around 62% for the five-stage classification. This highlights the advantage of fine-tuning HeartBERT's layers in capturing intricate ECG signal patterns, leading to better classification results.

We also expand our work to include heartbeat classification utilizing the Icentia11k dataset [31, 32]. Heartbeat classification is critical for detecting and monitoring arrhythmias and other cardiac conditions [33]. Our best results in this task show an F1 score of about 88%, further demonstrating the versatility and effectiveness of HeartBERT across various cardiovascular-related tasks.

To summarize, the contributions of this paper are as follows:
- Inspiring from the patterns within periodic time series data (i.e., natural language and ECG data) to bridge between NLP and the healthcare domain,
- Translating ECG signals into synthetic language representations using a quantization approach,
- Developing a versatile model that learns contextual embeddings from unlabeled ECG signals,
- Incorporating the proposed model into fine-tuning downstream tasks, namely sleep-stage classification and heartbeat classification, with various scenarios for updating the transformer layers.

The remainder of this paper is organized as follows: Section 2 reviews key related work and positions our contribution within the broader context of the field. Section 3 introduces the proposed HeartBERT model, providing details on tokenization and model architecture. Section 4 presents the experimental study, where we evaluate the model's performance on downstream tasks. Finally, Section 5 concludes the paper with a summary of our findings and suggestions for future research directions.

## 2. Related Work

From a technical point of view, the related research on the ECG domain could be grouped based on the learning methods employed to train the models into two essential classes: supervised approaches and unsupervised methods, which include self-supervised techniques. We first explore the supervised approaches and then the unsupervised methods. Although supervised approaches seem efficient, the lack of labeled data makes unsupervised learning, particularly self-supervised approaches, more prevalent.

### 2.1 Supervised approaches

Supervised learning approaches face several challenges. Firstly, they require a wealth of labeled data for effective training. Secondly, these models often struggle to generalize well to new and unseen datasets.

Many studies have employed traditional machine learning techniques for various ECG tasks. For instance, Mazaheri and Khodadadi [34] proposed a computer-aided diagnosis (CAD) system for classifying different types of cardiac arrhythmias using ECG signals. The DeepArr model, introduced in [35], serves as an investigative tool for arrhythmia detection, combining

1-D convolutional and Bi-LSTM layers to effectively learn features. Peng et al. [36] developed CS-TRANS, a deep learning framework regulated by stationary wavelet transform (SWT), which integrates CNN-SWT and a Transformer encoder for ECG denoising and classification. Mousavi et al. [37] proposed an ECG Language Processing method leveraging deep models like CNN and RNN for analyzing ECG signals but employed a simple approach for segmentation and vocabulary creation, leaving room for further refinement in future research. Additionally, a transformer-based deep neural network, ECG DETR, was presented in [38] for arrhythmia detection on continuous single-lead ECG segments, simultaneously predicting heartbeat positions and categories. Huang et al. [39] proposed a novel input representation that combines the RR interval with ECG signals to capture temporal differences in arrhythmias. Using a CNN-LSTM-Attention model, their method demonstrated the benefits of integrating periodic signal features in ECG analysis.

Akan et al. [40] explored the use of the Transformer architecture for heartbeat arrhythmia classification. The authors split a series of ECG waves into patches and turned them into embeddings. These sequences of embeddings were then fed into a standard Transformer encoder for classification purposes. However, the paper lacks detailed explanations of the patching and embedding processes, making their work difficult to reproduce. Another innovative model was proposed in [41]. This model considered both spatial and channel-wise features. Initially, the ECG signals were processed by a multi-scale convolutional component to capture spatial features at various scales using multiple convolutional layers, each with a distinct kernel size. Subsequently, the Channel Recalibration Module (CRM) adjusted the feature map to reflect distinctions between different leads. The resulting feature map served as the input to the third module, the Bidirectional Transformer (BiTrans). BiTrans extended the standard Transformer architecture by incorporating an inversion mechanism to include reversed sequence inputs, enabling the model to utilize both past and future ECG waveform contexts for a richer contextual understanding.

These transformer-based models require substantial labeled data to perform effectively, which imposes significant challenges. Acquiring annotated ECG data is labor-intensive and time-consuming, limiting practical application. Additionally, the need for large datasets demands considerable computational resources for training, further restricting their widespread use.

To achieve higher accuracy and better performance, some studies focus on manually crafting useful feature sets rather than relying on deep learning methods. For instance, Sarankumar et al. [42] employed Wavelet Transform (WT) to calculate an extensive set of features from 12-lead ECG signals, including time-domain, frequency-domain, and morphological features. These features were then used to classify the signals into nine categories: eight forms of arrhythmia and one normal, achieving an accuracy of 99.32%. While feature engineering can enhance performance, it requires domain expertise, and the extracted features are often task-specific, limiting their applicability to other tasks. In contrast, our proposed model is pre-trained once and can be employed for a range of downstream tasks with a minimal amount of labeled data.

## 2.2 Unsupervised (Self-supervised) approaches

Due to the limitations of supervised learning approaches, such as the need for large amounts of labeled data and limited generalizability, recent research has shifted towards unsupervised learning methods. In this context, self-supervised learning has gained significant attention due to its advantages over traditional unsupervised methods like clustering. Since the proposed model in this study also adopts a self-supervised learning approach, it is essential to review the latest work in this area. Self-supervised models typically involve two stages: pre-training on large datasets followed by fine-tuning for specific tasks or different datasets.

A prominent example is the model introduced by [43], which followed this two-step approach. The base model consisted of an encoder and a decoder and was pre-trained using a masked autoencoder (MAE) technique. In this method, parts of the input signal were removed before being passed to the encoder, while the decoder attempted to reconstruct the complete signal. This process enabled the model to learn the general features of ECG signals effectively, turning it into a robust feature extractor. In the fine-tuning stage, the decoder was removed, and a simple perceptron neural network was added to the model and was trained using a small set of labeled data for a specific task, such as arrhythmia classification. By leveraging the features learned during pre-training, the model achieved robust performance even on small, labeled datasets.

Mehari et al. [10] conducted a thorough evaluation of self-supervised learning for ECG data, utilizing advanced techniques such as contrastive learning. Their findings demonstrated that self-supervised models performed comparably to supervised ones, with improvements in label efficiency and noise robustness, marking a step forward in ECG representation learning. Qin et al. [44] proposed a model for general ECG classification that is applicable across different patients and supports multiple classes. The model was pretrained using both unsupervised and supervised methods, followed by fine-tuning with various strategies. They also developed a hierarchical classifier head to leverage label relationships for enhanced performance.

Yang et al. [45] presented an encoder-based Biosignal Transformer model that leveraged a cross-data learning method to embed biosignal values, including EEG and ECG, into dense real-valued vectors. The model was pre-trained on abundant biosignal data and subsequently fine-tuned for downstream tasks. Unlike our proposed method, which utilizes the Byte Pair Encoding (BPE) tokenizer for input processing, their work tokenized the input into fixed-length segments (usually 1 second) without considering high-frequency and meaningful tokens. While this simple token recognition method has some advantages, it also has notable disadvantages, such as missing high-frequency meaningful tokens and consequently increasing vocabulary size. In contrast, our proposed model, by employing the BPE tokenizer, not only considers high-frequency and meaningful tokens but also better preserves long-distance dependencies.

## 3. The Proposed Model (HeartBERT)

At the intersection of ECG analysis and natural language processing (NLP), our work captures the shared attributes between these domains. Both ECG signals and textual representations are characterized by periodic time series containing meaningful patterns. Given these similarities, we introduce HeartBERT, a novel model built upon the RoBERTa architecture [17], renowned for its encoder-only design.

We pre-train the HeartBERT model from scratch using curated datasets. Raw ECG signals are transformed into synthetic language through discretization and quantization, making them suitable for encoding into a language-like format. This derived textual data is then fed into the HeartBERT tokenizer. The tokenized data is processed through transformer layers to generate encoded embeddings. Following pre-training, the HeartBERT model is ready for deployment in downstream tasks. In this section, we explain the translation of ECG signals into synthetic language, the tokenization process, and the HeartBERT model structure. We also provide details on the self-supervised training approach and model configuration.

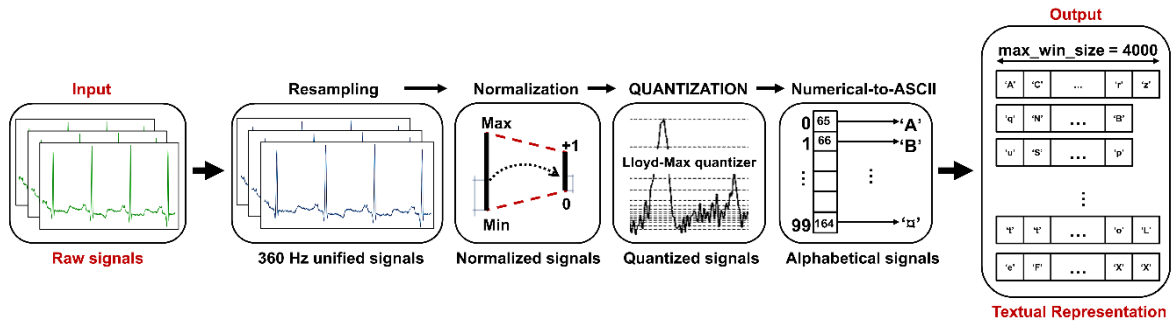

**Figure 1:** Illustration of signal conversion, including resampling, normalization, and quantization.

### 3.1 ECG Signals to Synthetic Language

Figure 1 illustrates the process of preparing data by converting raw ECG signals into synthetic language representations. This process involves several key steps:

- **Resampling:** All ECG signals undergo resampling to a frequency of 360 Hz using the Fast Fourier Transform (FFT) technique [46]. This resampling ensures uniformity in the data, making it suitable for subsequent processing steps[1] [47].
- **Normalization:** After resampling, the signals are normalized to a range of [0, 1]. This normalization step is crucial to standardize the data, eliminating variations due to differing amplitudes and facilitating consistent quantization.
- **Quantization:** To translate ECG signals into textual representations, we employ the Lloyd-Max quantization method [19, 48] with a quantization level set at 100. This method divides the continuous ECG signals into 100 discrete levels based on signal fluctuations, capturing detailed variations in the signal. Lloyd-Max quantization is a method typically used in signal processing to optimize the representation of a signal while minimizing distortion in the quantized signal [18].
- **Windowing:** Signal windowing is implemented with a maximum window size of 4000. This step segments the ECG signals into manageable chunks, making the data processing more efficient and ensuring that each window captures significant patterns within the signal.
- **Numerical-to-ASCII Conversion:** Each data point within a signal window is mapped to an ASCII character through a numerical-to-ASCII conversion function. This conversion transforms the quantized numerical values into characters, forming the basis of our synthetic language.

Once the data is converted into ASCII characters, it is ready to be fed into an off-the-shelf tokenizer. This tokenizer, discussed in Section 3.2, turns the data into tokens, with each token made up of characters from '**A**' to '¤' (100 letters of the alphabet) in our synthetic language.

---

[1] For resampling, we use the NeuroKit2 library, which is a comprehensive toolbox for physiological signal processing in Python (https://neuropsychology.github.io/NeuroKit/functions/signal.html).

## 3.2 Tokenization and Model Architecture

### 3.2.1 Tokenization

As shown in Figure 2(a), we start the pre-training process by training a SentencePiece BPE (Byte Pair Encoding) tokenizer[1]. BPE, known for its versatile encoding capabilities, efficiently encodes various data types beyond just traditional text. It operates by identifying and replacing frequently occurring byte pairs with a single unused byte, thereby optimizing the balance between computational efficiency and linguistic expressiveness [20, 49, 50].

The SentencePiece BPE tokenizer decomposes the textual representation of ECG data into smaller units based on their frequency in the training data. This tokenizer is specifically chosen for its ability to capture and segment the most frequent patterns within the data, making it highly effective for our needs. Figure 2(b) shows that after training the tokenizer, it is used as a token segmenter. This trained tokenizer takes a quantized ECG signal as input and returns the tokenized data segmented into tokens, which are then ready to be fed into the model. Therefore, these tokens act as meaningful inputs for the model but may not be easily interpretable by humans.

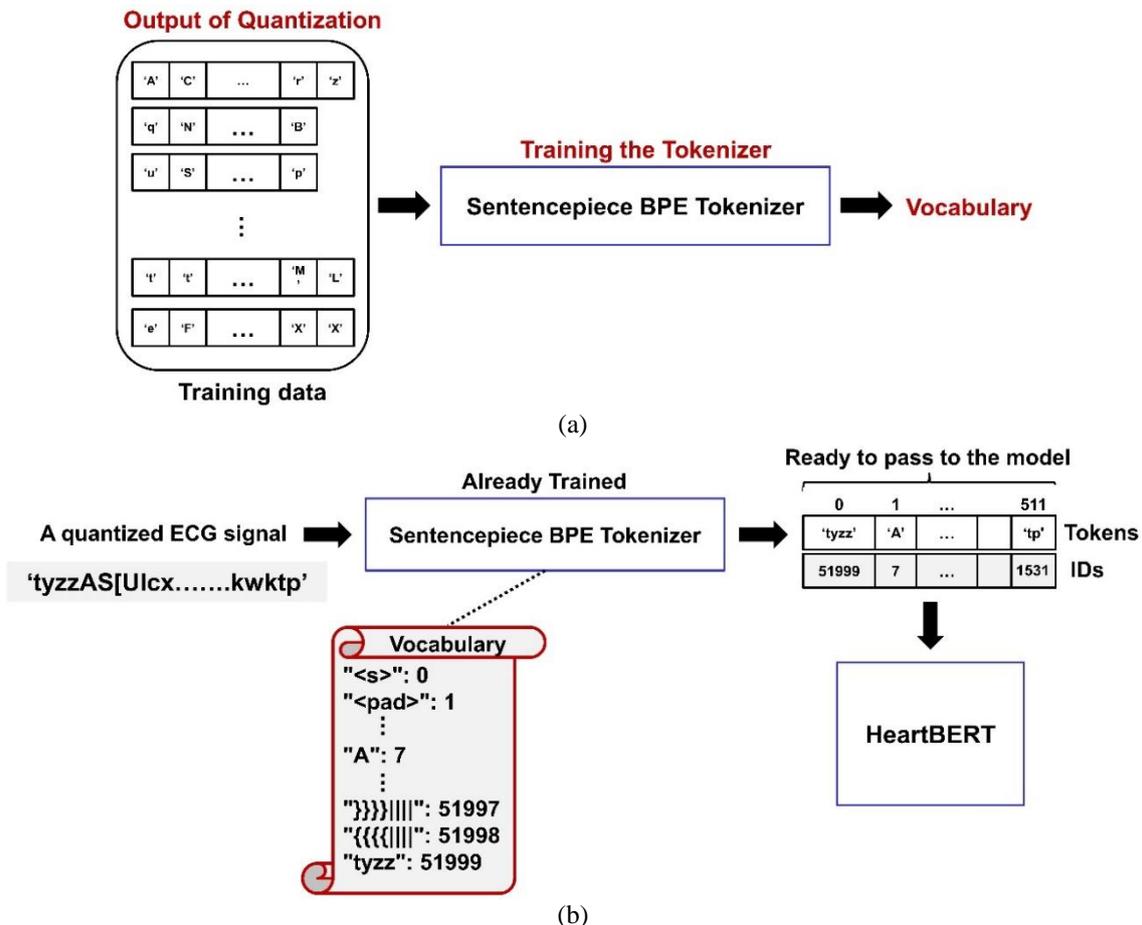

(a)

(b)

**Figure 2:** Training the Tokenizer. a) The vocabulary is obtained through tokenizer training. b) The trained tokenizer acts as a token segmenter.

Our tokenization approach offers significant advancements over traditional ECG tokenizers used in other works [45, 51, 52]. Conventional methods often rely on fixed-size windowing or simplistic segmentation, which may not effectively capture the nuanced patterns in ECG data. In contrast, the SentencePiece BPE tokenizer used in our model dynamically adjusts to the frequency of patterns, leading to a more flexible representation of the data. This flexibility is particularly beneficial in capturing both common and rare patterns within the ECG signals, thereby enhancing the model's ability to learn intricate cardiac nuances.

---
[1] We use the Hugging Face (https://huggingface.co/) libraries to implement our tokenizer and model.

### 3.2.2 HeartBERT Model

The HeartBERT model architecture, based on the RoBERTa framework [17], is designed to learn complex representations of ECG data. Figure 3 illustrates an overview of the model architecture. The key components are described as follows:

**Input Embeddings:** The input to the model consists of tokenized ECG data, where each token is embedded into a high-dimensional vector space. These embeddings are augmented with positional encodings to retain the order information of the tokens, which is essential for accurately representing the time-series nature of ECG data.

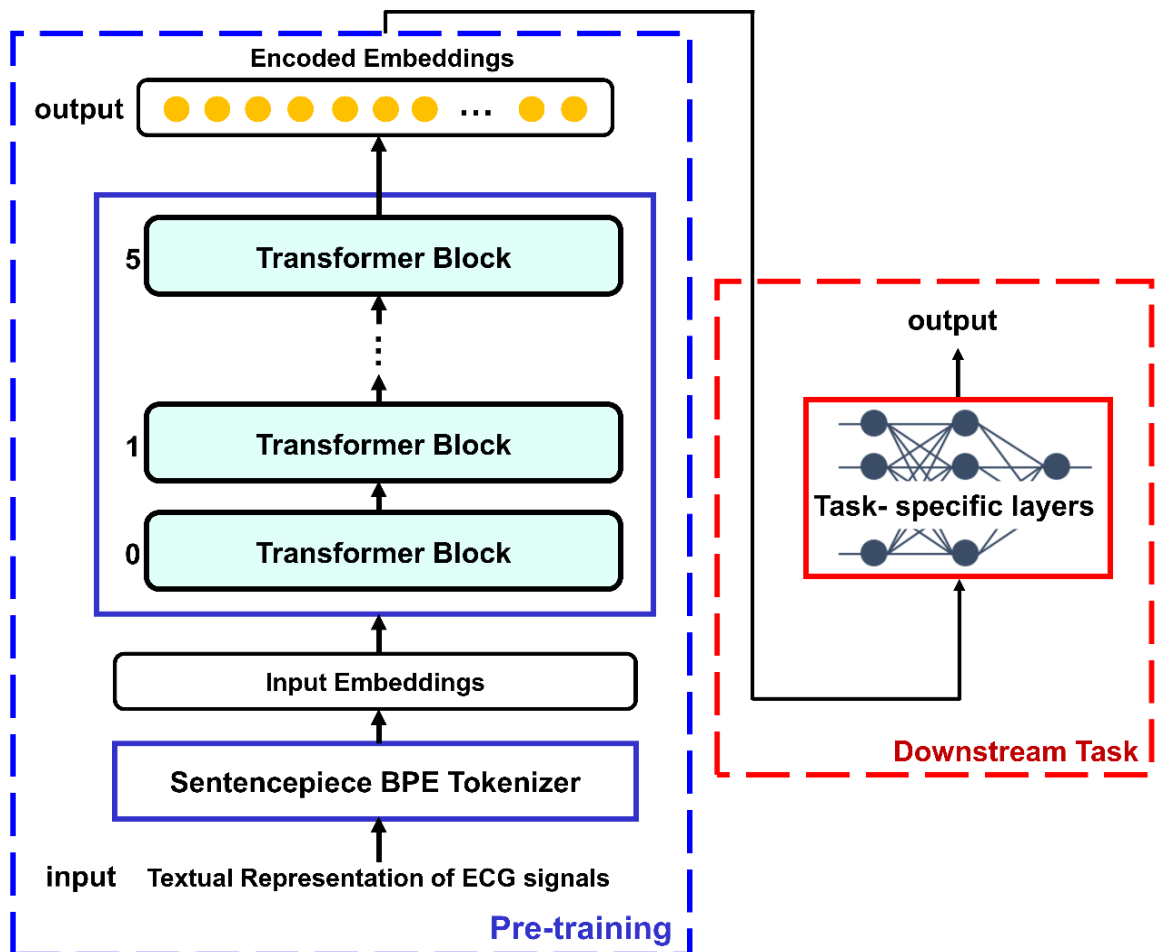

**Figure 3:** An overview of the HeartBERT model. The model is first pre-trained using textual representations of ECG signals as our training data, and subsequently utilized in fine-tuning for downstream tasks.

**Transformer Layers:** HeartBERT employs six transformer blocks, each comprising a multi-head self-attention mechanism [53] and a position-wise fully connected feed-forward network. The self-attention mechanism enables the model to focus on different parts of the input sequence, capturing dependencies and relationships within the ECG signals. This attention mechanism is crucial for identifying patterns that may span across different segments of the signal.

**Masked Language Modeling (MLM):** During pre-training, HeartBERT uses the MLM objective [54], where a certain percentage of tokens in the input sequence are randomly masked. The model is then trained to predict these masked tokens based on the surrounding context. This self-supervised learning approach helps the model capture the underlying structure and semantics of the ECG data without requiring labeled data.

Note that the architectural design of HeartBERT enables it to effectively learn contextual embeddings of ECG signals, which can be further employed in various downstream tasks.

### 3.3 Self-supervised Training

In this section, we delve into the self-supervised training process of HeartBERT, covering three key aspects: dataset selection, model configuration, and loss analysis.

#### 3.3.1 Datasets

For pre-training the HeartBERT model, we utilize three distinct datasets: MIT-BIH Arrhythmia Database [55, 56], PTB-XL [57], and European ST-T Database [58], selected for their relevance and breadth of data. Together, these datasets comprise over 314K rows, providing ample data for pre-training. Table 1 provides an overview of the statistical characteristics of these publicly available datasets, representing a diverse range of subjects and durations. As mentioned in Section 3.1, this data collection is the one that is preprocessed to create our translated ECG data in the form of a synthetic language. Remind that by employing self-supervised learning, the model learns to understand and represent intricate structures within ECG data without explicit labels or external supervision.

**Table 1:** Summary of pre-training datasets.

| Dataset | #Records | Rate | #Duration per record | #Subjects |
|---|---|---|---|---|
| MIT-BIH Arrhythmia Database [56] | 48 | 360 Hz | 30 minutes | 47<br>25 men, 22 women |
| PTB-XL [57] | 21837 | 500 Hz, 100 Hz | 10 seconds | 18885<br>(52% men, 48% women) |
| European ST-T Database [58] | 90 | 250 Hz | 2 hours | 79<br>(70 men, 8 women, one with no information) |

#### 3.3.2 Model Configuration and Settings

Following the two-stage training approach for HeartBERT—pre-training and downstream tasks— we detail the settings used during the pre-training phase here. The configuration for the downstream tasks is discussed in Section 4.3.

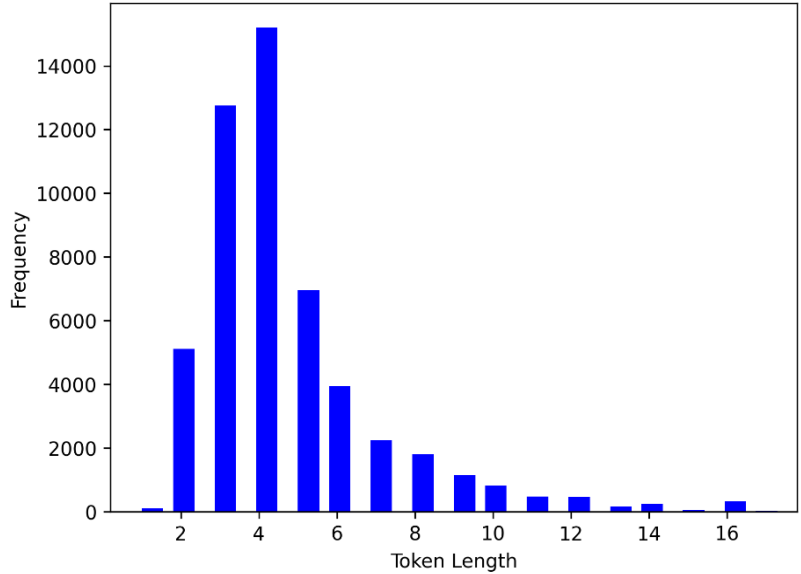

**Figure 4:** Histogram of token lengths. The low frequency of samples for lengths greater than 16 has been cropped.

**Table 2:** Configuration settings for HeartBERT pre-training phase.

| **Tokenization Parameters** | |
|---|---|
| Vocabulary Size | 52,000 tokens |
| Minimum Token Length | 1 |
| Maximum Token Length | 36 |
| Average Token Length | 4.66 |
| Maximum Sequence Length | 512 tokens |
| Probability of Masking | 15% |
| **Model Parameters** | |
| Number of Transformer Layers | 6 |
| Number of Attention Heads per Layers | 12 |
| Embedding size | 768 |
| Total Trainable Parameters | 83,504,416 |
| **Training Parameters** | |
| Batch Size | 64 |
| Learning Rate | 5e-5 |
| Optimizer | AdamW |
| Number of Epochs | 1000 |

Table 2 reports the configuration settings used in our experiments. We set the vocabulary size to 52,000, which is the total number of unique tokens the tokenizer learns. Figure 4 displays the histogram of token lengths, with the minimum token length being 1, the maximum being 36, and the average being 4.66. The maximum sequence length is set to 512, and a padding strategy is applied. The embedding size is also set to 768.

Additionally, the probability that a token is masked during the MLM is set to 15%. HeartBERT employs 6 transformer layers, each equipped with 12 attention heads, resulting in a total of 83,504,416 trainable parameters.

For the pre-training process, we use a batch size of 64 and train the model for 1000 epochs. The AdamW optimizer is also utilized with a learning rate of 5e-5.

### 3.3.3 Loss Analysis

To track the model's learning progress, the loss values are monitored during pre-training. A steady decrease in the loss value indicates that the model is successfully learning to predict the masked tokens in the input sequences. Figure 5 illustrates the loss curve over the training epochs, showing how the loss value changes as the model trains. Initially, the loss value is high, reflecting the model's random predictions. However, as training progresses, the loss value decreases, indicating that the model is improving its predictions based on the surrounding context.

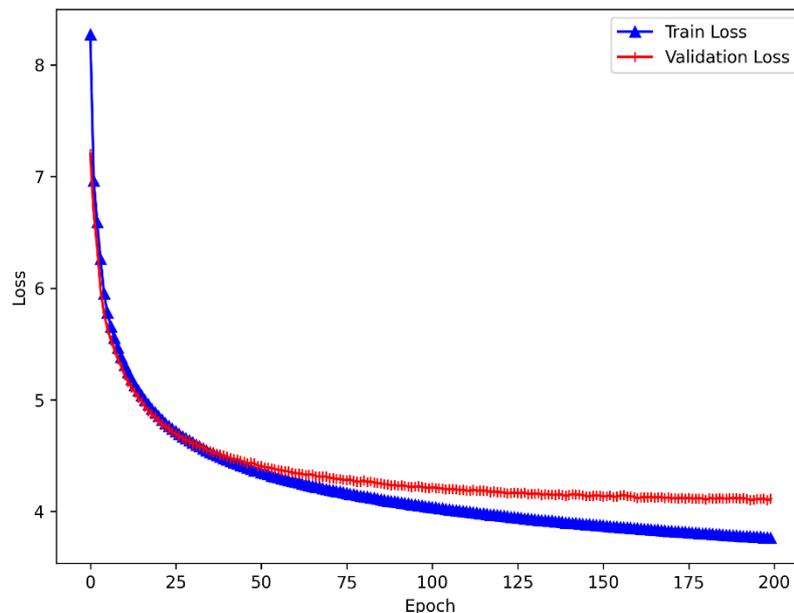

**Figure 5:** Trends of loss over epochs. The decreasing trend of loss indicates improvement in model performance with training iterations.

### 4. Experimental Study

Having completed the pre-training phase, we extract encoded embeddings enriched with contextual information, which serve as potent representations for downstream tasks. The sleep-stage and heartbeat classification tasks are two downstream applications that will serve to demonstrate the effectiveness and capabilities of the proposed model.

### 4.1 Downstream Tasks

**Task 1. Sleep-stage Classification:** During sleep, the body alternates through different physiological states called sleep stages, broadly categorized as non-rapid eye movement (NREM) and rapid eye movement (REM). NREM sleep comprises stages 1 (S1), 2 (S2), and 3 (S3), progressing in depth from light sleep to deep sleep. Stage 3 is characterized by slow brain waves and reduced muscle activity. REM sleep, on the other hand, is marked by rapid eye movements, increased brain activity, and muscle paralysis [59]. Accurate sleep-stage classification is crucial for understanding sleep patterns and providing insights into sleep-related health conditions. Disruptions in sleep stages can indicate disorders like insomnia or sleep apnea, while monitoring sleep stage transitions can aid in diagnosing and managing various health conditions linked to sleep disturbances, such as cardiovascular disease, depression, and cognitive decline [60]. In this study, our focus is on the automatic classification of ECG signal segments into sleep stages. We conduct both a three-stage classification, where S1, S2, and S3 are merged into a single NREM class, and a five-stage classification, which includes wake, REM, S1, S2, and S3.

**Task 2. Heartbeat classification:** ECG beat-by-beat analysis is essential for the early detection of some cardiovascular conditions. However, the variability in recording environments, differences in disease patterns among patients, and the complex, non-stationary, and noisy nature of ECG signals make heartbeat classification a challenging and time-consuming task for cardiologists [61]. Furthermore, the demand for automated ECG interpretation systems is steadily increasing with the growing use of devices like smartwatches that continuously monitor heart activity. Heartbeats consist of distinct waves, including the P wave, QRS complex, and T wave, which represent specific phases of the heart's electrical activity. These patterns can be disrupted by arrhythmias, leading to irregular heart rhythms where the heart may beat too slowly or too quickly. The Association for the Advancement of Medical Instrumentation (AAMI) standard [62] categorizes arrhythmias into five primary classes: Normal (N), Supraventricular abnormal (S), Ventricular abnormal (V), Fusion (F), and Unknown (Q).

**4.2 Data**

**Task 1:** For the **sleep-stage classification** task, we conduct experiments to fine-tune the HeartBERT model using the publicly available **MIT-BIH Polysomnographic Database (SLPDB)** [27]. This dataset, a resource of polysomnographic (PSG) recordings designed for sleep research, was originally annotated with six distinct sleep stages: wake, REM, S1, S2, S3, and S4. However, based on the standards introduced by the American Academy of Sleep Medicine (AASM) in 2007 [63], current studies typically consider only five stages: wake, S1, S2, S3, and REM. To align with this established framework, stages S3 and S4 are merged to represent the deep sleep stage.

The MIT-BIH Polysomnographic Database includes over 80 hours of recordings from 16 male subjects aged between 32 and 56 years (mean age 43), with weights ranging from 89 to 152 kg (mean weight 119 kg), many of whom suffer from sleep apnea. Each subject's data comprises four-, six-, or seven-channel recordings at a sampling rate of 250 Hz, encompassing ECG, EEG, EOG, EMG, respiratory effort, and oximetry signals.

**Table 3:** Dataset statistics for sleep-stage and heartbeat classification tasks.

| | MIT-BIH Polysomnographic (Sleep-stage Classification) | | | | | | Icentia11k (Heartbeat classification) | | | | |
|---|---|---|---|---|---|---|---|---|---|---|---|
| | #Wake | #REM | #S1 | #S2 | #S3 | #Total (after under-sampling) | #N | #S | #V | #Q | #Total (after under-sampling) |
| **Three-stage** | 31,030 | 7,000 | 18,140 | 38,830 | 6,630 | 33,150 | 2,061,141,216 | 19,346,728 | 17,203,041 | 676,364,002 | 20,000 |
| **Five-stage** | 31,030 | 7,000 | | 63,600 | | 21,000 | | | | | |

**REM**: Rapid Eye Movement, **N**: Normal, **S**: Supraventricular abnormal, **V**: Ventricular abnormal, **Q**: Unknown

Focusing exclusively on ECG data from this dataset, we resample records to 360 Hz to align with the pre-training phase and subsequently normalize them. Each 30-second annotated data segment is divided into 3-second segments, with corresponding labels preserved. To address class imbalance, an equal number of segments are selected from each sleep stage, with detailed statistics reported in Table 3. There are 21,000 preprocessed records (i.e., 3-second segments) for the three-stage classification and 33,150 preprocessed records for the five-stage classification in total. Moreover, the data is partitioned into training, validation, and test sets with a split of 70% for training, 10% for validation, and 20% for testing.

**Task 2:** One of the most commonly used datasets for **heartbeat classification** is the MIT-BIH dataset [56], and a significant portion of the results reported in the literature are based on this dataset. However, since this data was used in the pre-training phase of HeartBERT, it cannot be used for this experiment. Instead, to evaluate the model's effectiveness, the **Icentia11k** dataset is employed [31, 32]. This dataset contains continuous ECG signals from 11,000 patients, with over 2 billion labeled heartbeats. The signals were recorded using a single-lead chest-mounted device at a 16-bit resolution and a sampling frequency of 250 Hz, for durations of up to two weeks. The average age of the patients was $62.2 \pm 17.4$ years, and each heartbeat was reviewed and labeled for type and rhythm by 20 specialized technicians. It is worth mentioning that this dataset classifies heartbeats into only four categories: Normal (N), Supraventricular abnormal (S), Ventricular abnormal (V), and Unknown (Q). The Fusion beat (F) class is not included.

Just like in our previous dataset used for the sleep task, due to the discrepancy in sampling rates between the signals in this dataset and those used during the pre-training phase of the main model, it is necessary to resample the signals to match the pre-training rate of 360 Hz at the beginning of the process. After resampling, the ECG recordings are segmented into consecutive heartbeats. Since the R-peak locations are already provided in the dataset, the only required step is to define the midpoint between two R-peaks as the endpoint of the previous heartbeat and the starting point of the next.

One of the primary objectives of utilizing the pre-trained model is to significantly reduce the need for large amounts of labeled data. To evaluate the impact of HeartBERT, only 20,000 heartbeats are selected from a total of over two billion, as shown in Table 3. It is important to note that medical datasets are typically highly imbalanced, and various techniques are used to address this issue. For instance, some studies adjust the cost function to prevent class imbalance from affecting classification accuracy [41, 64], while others apply oversampling or under-sampling techniques [65]. Since we use a small subset of the total data, we have the flexibility to select an equal number of heartbeats from each class. As a result, the final dataset is fully balanced, containing 5,000 heartbeats from each class.

### 4.3 Configurations

Here, we outline the implementation details of our model during fine-tuning and describe the baseline architecture for comparison. Moreover, all our experiments are performed on a single T4 GPU with 15 GB RAM using the PyTorch framework.

As illustrated in Figure 6, we employ a hybrid architecture that combines the pre-trained HeartBERT model with a bidirectional Long Short-Term Memory (Bi-LSTM) network for both sleep-stage and heartbeat classification tasks.

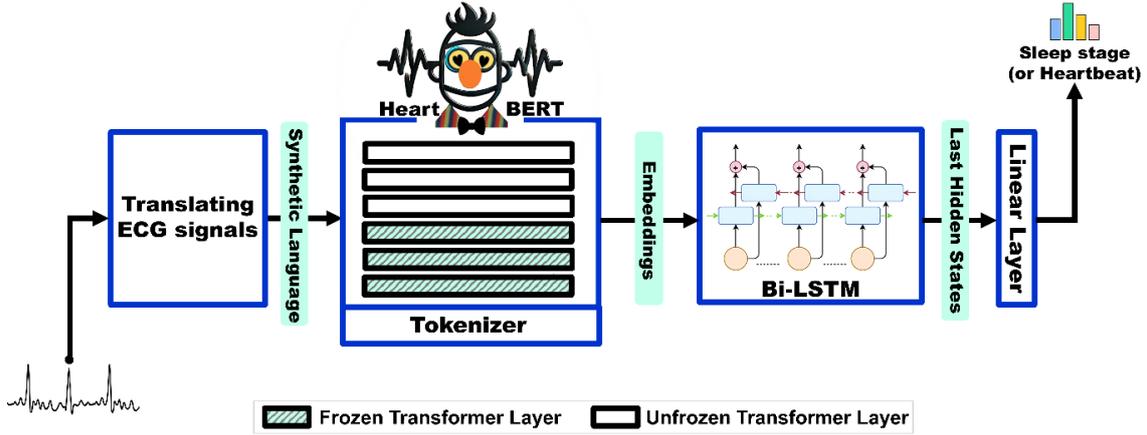

**Figure 6:** The hybrid architecture of our model combining HeartBERT and Bi-LSTM for two downstream tasks—sleep-stage and heartbeat classification. We evaluate different configurations by freezing various layers of HeartBERT.

**Table 4:** The number of trainable parameters in different configurations of the Hybrid Proposed Model (HeartBERT + Bi-LSTM Head). The variation in the number of trainable parameters across different classification tasks is minimal due to the consistent model architecture. Adding additional output classes has little impact on the overall parameter count.

| Model | #Trainable Parameters (3-stage sleep classification) | #Trainable Parameters (5-stage sleep classification) | #Trainable Parameters (Heartbeat classification) |
|---|---|---|---|
| HeartBERT (all-frozen) + Bi-LSTM Head | 1,510,915 | 1,511,429 | 1,511,172 |
| HeartBERT (1-unfrozen) + Bi-LSTM Head | 8,598,787 | 8,599,301 | 8,599,044 |
| HeartBERT (half-frozen) + Bi-LSTM Head | 22,774,531 | 22,775,045 | 22,774,788 |
| HeartBERT (all-unfrozen) + Bi-LSTM Head | 44,038,147 | 44,038,661 | 44,038,404 |

**Bi-LSTM Layer:** Following the HeartBERT embedding layers, a Bi-LSTM layer is incorporated to capture sequential dependencies within the extracted features. The Bi-LSTM is configured with an input size of 768 (corresponding to the output size of the HeartBERT embeddings) and a hidden size of 128. This layer is bidirectional to enhance the model's capability to capture both forward and backward context information in an input ECG sequence.

**Classification Head:** A fully connected linear layer follows the Bi-LSTM output to map the final hidden states to the number of output classes.

For training the model, we employ the Cross-Entropy loss, suitable for multi-class classification tasks, to measure the discrepancy between predicted and actual labels. The model parameters are optimized using the Adam optimizer. We experimented with multiple learning rates, specifically 3e-5, 4e-3, and 5e-3, and selected the one that yielded the best performance. The training process is carried out with a batch size of 8.

**Integration of HeartBERT:** The HeartBERT model serves as the feature extractor from ECG signals. During the fine-tuning process, we experiment with freezing different numbers of layers of the HeartBERT model to investigate their impact on performance. Our model's scalability covers a spectrum of configurations, ranging from approximately 1.5M trainable parameters to over 44M parameters, as detailed in Table 4.

### 4.4 Counterparts

**Task 1. Sleep-stage classification:** To evaluate our model's performance, we compare our proposed model with a deep convolutional recurrent (DCR) model proposed by [66] due to its methodological similarity. Both models utilize ECG signals for **sleep-stage classification** without relying on handcrafted features or extracting any intermediate signs like the RR interval and HR. This approach allows the models to learn directly from the raw ECG data, capturing more nuanced and complex patterns. The DCR model is advantageous as it effectively captures both local and sequential patterns in the data, making it a reliable choice for this task. It employs a convolutional neural network (CNN) with three layers of one-dimensional (1D) convolutions and 1D pooling to capture local patterns within the ECG sequences. The extracted features are then processed by two gated recurrent unit (GRU) layers, which are designed to capture sequential patterns. Finally, a fully connected layer classifies the features into different sleep stages. The DCR model also leverages batch normalization, dropout, and rectified linear unit (ReLU) activation for optimization.

**Task 2. Heartbeat classification:** To evaluate the performance of HeartBERT, the Multimodal Image Fusion (MIF) model, presented in [65], is chosen as the baseline. Both HeartBERT and MIF are applied to the same dataset, and their performance is compared in terms of heartbeat classification accuracy.

The MIF model converts ECG signals into three distinct types of images: Gramian Angular Fields (GAF), Recurrence Plots (RP), and Markov Transition Fields (MTF). These transformations capture different statistical methods for representing the temporal dynamics of the ECG signal. Each of these image types provides unique insights into the signal's structure. These images are then combined into a three-channel image, similar to an RGB image, and fed into the AlexNet convolutional neural network to extract features and classify heartbeats. This multimodal approach enhances classification accuracy and is potentially valuable in clinical applications.

### 4.5 Results and Analysis

#### 4.5.1 Sleep-stage classification

This section presents the performance comparison of various models on the MIT-BIH Polysomnographic Database [27] for both three-stage and five-stage classification tasks. The evaluation metrics, including precision, recall, F1-score, and accuracy, are reported in Table 5. We explore different model configurations by adjusting the updating layers of HeartBERT during fine-tuning to analyze how they impact classification performance.

**Three-Stage vs. Five-Stage Sleep Classifications**: In general, we observe that the results for the three-stage classification are better than those for the five-stage classification. This disparity is expected, as the three-stage classification groups similar sleep stages together, reducing complexity and improving performance. In contrast, the five-stage classification requires distinguishing between more subtle differences in the data, making it inherently more challenging and resulting in lower performance metrics.

**Impact of Embeddings**: Providing raw ECG input to the DCR model, compared to the embeddings provided by our proposed HeartBERT model without fine-tuning its layers, clearly shows the effect of our model in capturing ECG patterns in the contextual representations. The HeartBERT model, even with all layers frozen, significantly enhances the performance metrics. This implies that the pre-trained HeartBERT embeddings effectively capture and represent the essential patterns in the ECG data, leading to improved classification performance. These embeddings also enable the model to perform the fine-tuning phase on the downstream task with much less annotated data, highlighting one of the most advantageous aspects of our work.

**Table 5:** Performance comparison of different models on the MIT-BIH Polysomnographic Database for three- and five-stage classification tasks. The best accuracy results are shown in **bold**.

| Model | | Three-stage classification | | | | Five-stage classification | | | |
|---|---|---|---|---|---|---|---|---|---|
| | | Precision | Recall | F1 | Accuracy | Precision | Recall | F1 | Accuracy |
| DCR Model [66] | Micro | 0.3724 | 0.3724 | 0.3724 | 0.3724 | 0.2448 | 0.2448 | 0.2448 | 0.2448 |
| | Macro | 0.3818 | 0.3687 | 0.3459 | | 0.2443 | 0.2442 | 0.2132 | |
| HeartBERT (all-frozen) + DCR Model Head | Micro | 0.6652 | 0.6652 | 0.6652 | 0.6652 | 0.5142 | 0.5142 | 0.5142 | 0.5142 |
| | Macro | 0.664 | 0.6644 | 0.663 | | 0.5231 | 0.5135 | 0.5092 | |
| HeartBERT (all-frozen) + Bi-LSTM Head | Micro | 0.6240 | 0.6240 | 0.6240 | 0.6240 | 0.5240 | 0.5240 | 0.5240 | 0.5240 |
| | Macro | 0.6209 | 0.6238 | 0.6203 | | 0.5229 | 0.5234 | 0.5182 | |
| HeartBERT (1-unfrozen) + Bi-LSTM Head | Micro | 0.7464 | 0.7464 | 0.7464 | **0.7464** | 0.6237 | 0.6237 | 0.6237 | **0.6237** |
| | Macro | 0.7459 | 0.7465 | 0.7462 | | 0.6184 | 0.6138 | 0.6141 | |
| HeartBERT (half-frozen) + Bi-LSTM Head | Micro | 0.6869 | 0.6869 | 0.6869 | 0.6869 | 0.6143 | 0.6143 | 0.6143 | 0.6143 |
| | Macro | 0.6869 | 0.6877 | 0.6842 | | 0.6184 | 0.6138 | 0.6141 | |
| HeartBERT (all-unfrozen) + Bi-LSTM Head | Micro | 0.6440 | 0.6440 | 0.6440 | 0.6440 | 0.5964 | 0.5964 | 0.5964 | 0.5964 |
| | Macro | 0.6579 | 0.6447 | 0.6465 | | 0.5961 | 0.5963 | 0.5960 | |

**Impact of HeartBERT Model Layers during Fine-Tuning**: Unfreezing the HeartBERT layers during fine-tuning shows varying impacts. The one-layer unfrozen configuration achieves the best performance with 0.7464 accuracy for three-stage classification and 0.6237 accuracy for five-stage classification. This is followed by the half-layer unfrozen configuration, and finally the all-layer unfrozen configuration.

The observed decline in performance with increasing numbers of unfrozen layers can be justified by the nature of pre-trained language models. When too many layers are unfrozen, the model may overfit to the specific task, losing the generalizability provided by the pre-trained representations. The optimal performance with one layer unfrozen suggests that minimal fine-tuning allows the model to leverage the robust features learned during pre-training while adapting sufficiently to the specific downstream task, thereby reducing the computational demands and hardware requirements needed to run experiments, which is a practical advantage of our model.

Figure 7 shows the sleep-stage classification results per class, clearly distinguishing the model's performance across different sleep stages. In Figure 7(a), the model demonstrates strong performance in classifying the Wake and REM stages, with balanced precision and recall values, particularly excelling in REM classification. This suggests that the model effectively captures the physiological patterns of rapid eye movement (REM) sleep, likely reflected in heart rate variability, which helps distinguish REM from other stages in the ECG signal. Although still solid, NREM classification lags slightly behind, likely due to blurred boundaries between stages, which affects both precision and recall.

In Figure 7(b), when classification is extended to finer-grained sleep stages, a notable drop in performance is observed for lighter stages like S1 and S2, where physiological signals overlap more significantly, making these stages harder to distinguish. This is reflected in the relatively lower F1 scores, particularly for S2, highlighting the challenges of differentiating the subtle variations of early non-REM sleep. In contrast, S3, representing deep sleep, is classified much more accurately, likely due to the more pronounced physiological markers associated with this stage. REM, though slightly less well-identified than in Figure 7(a), remains robust, reaffirming the model's ability to detect the distinct features of REM sleep despite the increased complexity of finer classification.

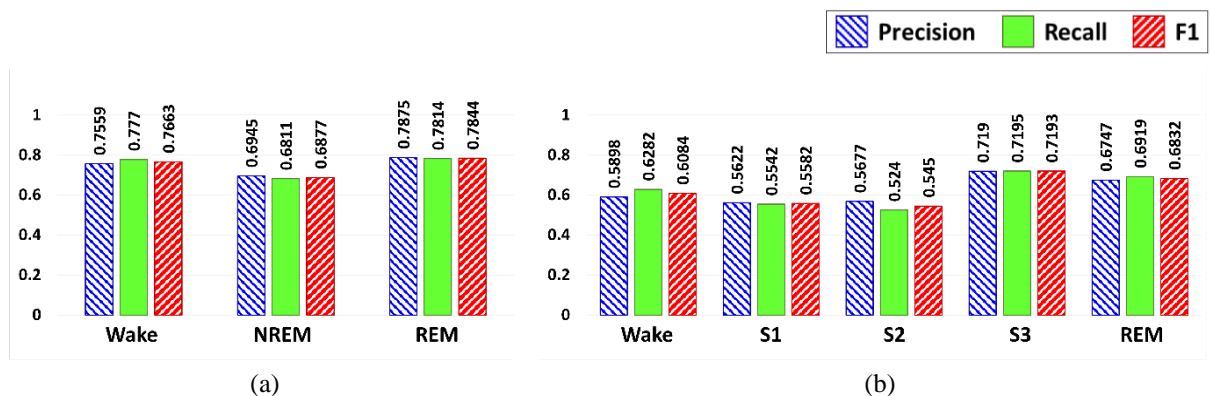

**Figure 7:** Class-wise performance metrics (Precision, Recall, and F1) for sleep-stage classification. a) Three-stage classification, b) Five-stage classification.

### 4.5.2 Heartbeat classification

To evaluate the proposed model's performance in heartbeat classification, we first examine the effect of fine-tuning different numbers of HeartBERT layers. Based on these results, the best-performing model is selected for the second experiment, in which it is compared with the baseline Multimodal Image Fusion (MIF) model [65]. The first experiment's results, presented in Table 6, indicate that optimal performance occurs when the last three HeartBERT layers are fine-tuned. Consequently, this configuration is chosen for further comparison.

In the second experiment, the MIF model serves as a baseline. The results are presented in Table 7. Due to the high computational cost of the MIF model's transformations[1], it is not feasible to use the entire dataset. Instead, 2,000 randomly selected samples from each class are used, resulting in a total of 8,000 labeled heartbeats for evaluation.

---

[1] A significant performance gap is observed between our implementation of the MIF model and the original MIF model reported in the literature. While the original MIF model paper has reported an accuracy of approximately 98% for heartbeat classification on the MIT-BIH dataset [58], our experiment with the MIF model yields poor results due to the limited training data. The MIF model was trained on 152,471 heartbeats, whereas our experiment uses only 8,000, about 20 times fewer data points.

**Table 6:** Performance comparison of different models for the heartbeat classification task using the Icentia11k dataset. The best accuracy is in **bold**.

| Model | | Precision | Recall | F1 | Accuracy |
|---|---|---|---|---|---|
| HeartBERT (all-frozen) + Bi-LSTM Head | Micro | 0.8535 | 0.8535 | 0.8535 | 0.8535 |
| | Macro | 0.8592 | 0.8543 | 0.8547 | |
| HeartBERT (1-unfrozen) + Bi-LSTM Head | Micro | 0.8484 | 0.8484 | 0.8484 | 0.8484 |
| | Macro | 0.8529 | 0.8502 | 0.8480 | |
| HeartBERT (half-frozen) + Bi-LSTM Head | Micro | 0.8848 | 0.8848 | 0.8848 | **0.8848** |
| | Macro | 0.8854 | 0.8850 | 0.8852 | |

Despite this fact, our proposed model demonstrates remarkable robustness and adaptability, even with a significantly smaller dataset. Unlike the MIF model, which shows a noticeable drop in performance when generalizing to new data, our model efficiently captures key features with fewer labeled samples. This highlights one of the key strengths of our model—its ability to maintain competitive performance with less computational resources and data, making it more practical and scalable for real-world applications, particularly in low-resource settings.

**Table 7:** Performance comparison between our proposed model and the baseline for the heartbeat classification task using the Icentia11k dataset.

| Model | | Precision | Recall | F1 | Accuracy |
|---|---|---|---|---|---|
| Multimodal Image Fusion (MIF) [65] | Micro | 0.332 | 0.332 | 0.332 | 0.332 |
| | Macro | 0.350 | 0.340 | 0.260 | |
| HeartBERT (half-frozen) + Bi-LSTM Head | Micro | 0.8051 | 0.8051 | 0.8051 | **0.8051** |
| | Macro | 0.8319 | 0.8054 | 0.8095 | |

The class-wise performance metrics for heartbeat classification in Figure 8 show strong results across Normal (N), Unknown (Q), Supraventricular abnormal (S), and Ventricular abnormal (V) categories. The model performs best with class V (ventricular abnormalities), achieving an F1 score of 0.9661, likely due to distinct physiological markers. Class S (supraventricular abnormalities) also shows high performance, with an F1 score of 0.9075, reflecting the model's effectiveness in detecting these signals despite their subtler features. For Normal heartbeats (class N), the F1 score of 0.8350 indicates reliable performance, though there may be minor overlap with abnormal classes. The classification of Unknown beats (class Q) yields an F1 score of 0.8322, reflecting the challenge of distinguishing this heterogeneous group. Overall, the model excels in identifying abnormal heartbeats while maintaining solid performance for normal and unknown categories.

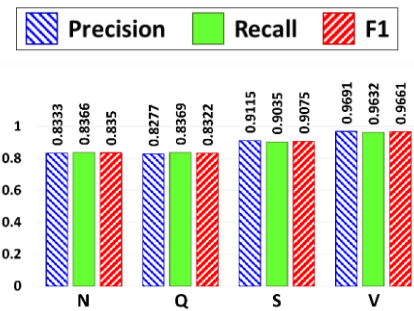

**Figure 8:** Class-wise performance metrics (Precision, Recall, and F1) for heartbeat classification across four categories: Normal (N), Unknown (Q), Supraventricular abnormal (S), and Ventricular abnormal (V).

## 5. Conclusion

In this work, we presented **HeartBERT**, a novel model that applies self-supervised representation learning inspired by NLP advancements to analyze ECG signals. By leveraging the RoBERTa architecture and transforming ECG signals into a synthetic language, HeartBERT bridges the gap between natural language processing and healthcare, offering a robust framework for analyzing periodic time-series data. The results from our experiments on sleep-stage classification and heartbeat classification demonstrated the model's efficacy in extracting meaningful patterns, achieving state-of-the-art performance with limited labeled data. While HeartBERT sets a strong foundation for self-supervised ECG analysis, there are several promising avenues for future work. Expanding HeartBERT to multi-channel modeling of ECG signals could improve its ability to capture complex inter-lead relationships, further enhancing its performance. Additionally, extending this framework to other biosignals, such as electroencephalogram (EEG), could unlock new possibilities for modeling and interpreting physiological data in domains beyond cardiology.

This paper is currently under investigation/review, and further changes may be applied in the future.